\documentclass[10pt]{article}
\usepackage{graphicx}
\usepackage{amsmath}
\usepackage{amsfonts}
\usepackage{amssymb}

\def\d{{\rm d}}\def\p{\partial}\def\re{\textrm{Re}\,}
\newcommand{\ang}[1]{\stackrel{\baselineskip=0pt\frown }{#1}}
\title{Outline of the Unified Theory of Spiral \\ and Bar-like Structures in Galaxies}
\author{E. V. Polyachenko\thanks{E-mail: epolyach@inasan.rssi.ru} \\
       Institute of Astronomy, Moscow 119017, Russia}
\date{}

\begin{document}

\maketitle

\label{firstpage}

\begin{abstract}
This paper presents a new approach to studying galactic structures.
They are considered as the
low-frequency normal modes in a disc of orbits precessing at different
angular speeds. Such a concept is an adequate
alternative to the commonly used approach of treating the disc as a set of
individual stars rotating at near-circular orbits around the centre.
The problem of determining the normal modes is reduced to a simple
integral equation in the form of the classical eigen-value problem, where the
eigen-value is directly equal to the pattern speed of the mode, $\Omega_p$.
An examination of the general properties of the basic integral equation shows
that two types of solutions exist, bar-like and spiral.
The numerical solutions of both types are obtained.
The characteristic pattern speeds are of the order of the mean orbit
precession speed, although for the bar-modes $\Omega_p$ can markedly exceed
the maximum precessing speed of orbits. 
It is shown that the bar-mode grows due to the immediate
action of its gravitational field on the stars at the resonance regions.
As for the spiral mode, its excitation is probably due to the inner Lindblad
resonance that can promote mode growth. 
\end{abstract}

\vspace{1mm}\noindent
Keywords: Galaxies: structure.

\section{Introduction}
A disc galaxy is primarily a set of stars which rotate around the centre
at near-circular orbits at the angular velocity $\Omega(r)$. The
observed spiral and bar-like structures are customarily treated as
perturbations in such a differentially rotating disc. Thus, one might expect
that the typical pattern speeds of these structures, $\Omega_p$,
should be of
the order of some average star angular velocity, $\bar \Omega$.
In reality, however, the pattern speeds $\Omega_p$ are of the order of the
characteristic precession speed of star orbits, $\bar\Omega_{pr}$, which is
only a small fraction of $\bar \Omega$. Recall that $\Omega_{pr} =
\Omega(r)-\kappa(r)/2$ for near-circular orbits, where $\kappa(r)=
(4\Omega^2+\d\Omega^2/\d r)^{1/2}$ is the epicyclic frequency.

According to Lynden-Bell (1979), if
stars involved in the formation of the structures
satisfy the inequality
\begin{equation} 
\epsilon \equiv |\Omega_p - \Omega_{pr}|/\Omega
\ll 1,
\label{in1}
\end{equation}
then each star orbit as a whole, but
not the individual stars, participates in the perturbations\footnote{Note, 
after Arnold (1989), that it was Gauss who had proposed,
  for studying the perturbations of planets by each other, to smear
  out a mass of each planet along its orbit in proportion to a time
  and replace the attraction of planets by the attraction of such
  rings.}.
Consequently, for studying such structures, it is more reasonable to
use the model of the disc of star orbits precessing with different
speeds than the commonly used concept of the differentially rotating
disc of individual stars. The main conjecture of the present paper is that
the galactic spirals and bars are the normal modes in such a model of
the disc galaxy. In view of its simplicity, this approach, as we will
see below, allows one to clarify the underlying physical mechanisms in
the formation of galactic structures.

The Lynden-Bell inequality (\ref{in1}) mentioned above are justified 
by results of numerous calculations (e.g., Athanassoula \& Sellwood (1986) 
for bar-modes or Lin, Yuan \& Shu (1969) for spirals).

\begin{figure}
\begin{center}
\includegraphics[width=14cm]{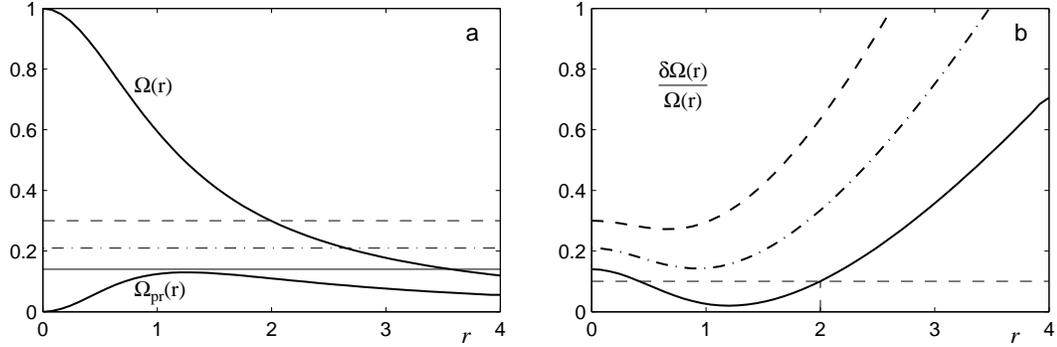}
\end{center}
\caption[]{%
Justification of the Lynden-Bell model of precessing orbits for
bar-modes studied by Athanassoula \& Sellwood (1986): a) the curves
$\Omega(r)$ and $\Omega_{pr}(r)$ for the Plummer potential (thick
solid lines), and the minimum, mean, and maximum patterns speeds (thin 
solid, dash-dotted, and dashed lines, respectively) from the list of
Athanassoula \& Sellwood (see their Table 1); b) the ratios
$\delta\Omega/\Omega$ for $\Omega_p$ from~a).} 
\label{fig_shust}
\end{figure}

Fig.\,\ref{fig_shust}a shows $\Omega(r)$ and $\Omega_{pr}(r)$ for the
Plummer potential $\Phi_0(r) = -(1+r^2)^{-1/2}$ used by  Athanassoula
\& Sellwood (1986) in their $N$-body bar-mode analysis. The solid thin
horizontal line $\Omega_p^{\min}=0.14$, the dash-dotted line
$\Omega_p=0.21$, and  the dashed line $\Omega_p^{\max} =0.3$
correspond, respectively, to the minimum, mean, and maximum pattern
speeds from the list given by Athanassoula \& Sellwood (1986) in their Table 1.
Fig.\,\ref{fig_shust}b represents the ratios $\delta\Omega/\Omega$
($\delta\Omega \equiv |\Omega_p - \Omega_{pr}|$) for these modes. For
the first mode localized inside the circle $r = 2$ (thin dashed lines 
in Fig. \,\ref{fig_shust}b), typical ratios are  of the order of 0.1.
For the second mode (mean pattern speed), these ratios are typically
of the order of 0.2. Even for the fastest mode ($\Omega_p =
\Omega_p^{\max}$), typically, $\delta\Omega/\Omega \sim 0.3$, taking
into account that this mode is more concentrated to the centre.
Therefore it is little wonder that the pattern speeds calculated
below (Section 5) in the framework of our approach repeat very
accurately those of Athanassoula \& Sellwood (1986) in all cases,
including the most rapid bar-modes.

\begin{figure}
\begin{center}
\includegraphics[width=14cm]{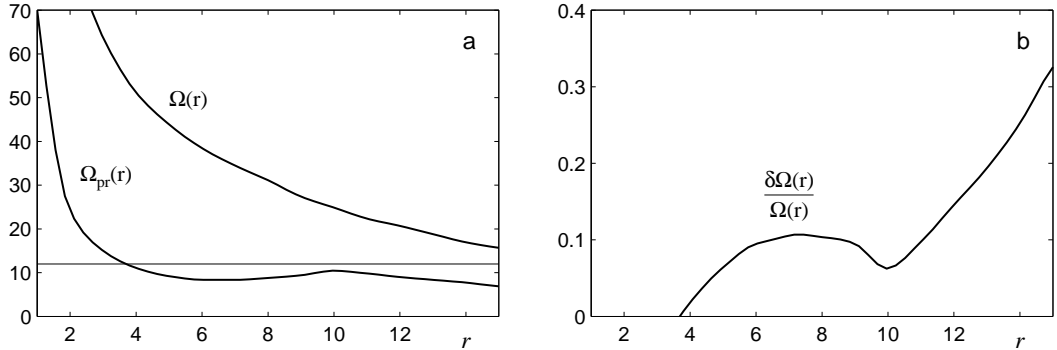}
\end{center}
\caption[]{%
The same as in Fig.\,\ref{fig_shust}, for the model of the Galaxy 
from Lin, Yuan, \& Shu (1969).}
\label{fig_shmidt}
\end{figure}

On the other hand, it is evident from the curves in
Fig. \,\ref{fig_shust}b that the inequality (\ref{in1}) and hence our
approach fail at sufficiently large radii.
This is certainly of no importance for determining the pattern speeds,
since the bar-modes are mainly localized near the centre.
Roughly speaking, the ratio $\delta\Omega/\Omega$ becomes of the order
of 1 near the corotation radius (i.e., at $r_c \approx 3.6, 2.7$, and
$2.0$, for the first, second, and third modes, respectively).

Fig.\,\ref{fig_shmidt}a,b show similar graphs for the
classical spiral example from Lin, Yuan \& Shu (1969) where the model
of our Galaxy was considered. The pattern speed used here corresponds to
the mean of the interval ($\Omega_p = 11 - 13$km s$^{-1}$ kpc$^{-1}$) suggested in
the cited paper, i.e., the horizontal line in Fig.\,\ref{fig_shmidt}a
is $\Omega_p=12$km s$^{-1}$ kpc$^{-1}$. 
Fig.\,\ref{fig_shmidt}b shows that the analysis of the eigen-modes of
Galactic disc for this model could be performed in the framework of 
our approach (without the restrictions of the WKBJ theory).
Note that substantially higher values of $\Omega_p$, compared to those
from Lin et al. (see, e.g.,
Blitz 1983), are most likely due to an independent peripheral stage of the
spiral structure. Such structures can be analyzed using the integral
equations described in the end of Section 2.

We picture the disc as comprising a large number of slowly deforming and
rotating elliptical rings, along which an individual stars moves so fast it
is no more than a blur.  Small changes with radius in the shape or
orientation of rings give rise to regions of enhanced surface density (see, 
e.g., Fig 6-11 of Binney \& Tremaine 1987). In the absence of self gravity any initial pattern of
over-densities will wind up  because the part of the pattern at radius $r$
will pecess at a rate $\Omega_{pr}$ that varies significantly with $r$.

Rings that have inclined minor axes exert gravitational torques on one
another. These torques cause the angular momenta and radial actions of the
rings slowly to change, thus deforming the rings and altering any initial
pattern of surface density.  The torques also affect the precession rates of
the ring's the minor axes. The
interaction between rings is strongest when the rings are physically
adjacent, i.e., have similar values of the actions. If we neglect the
non-negligible range of the interaction, the disc has something in common
with a goods train of deformable wagons connected by springs.  Self gravity
allows waves of ring deformation to propagate through the disc just as the
springs permit waves of wagon displacement and deformation to propagate 
down
the train. We show that these waves can generate a pattern that rotates
rigidly at an angular frequency that is larger than the fastest precession
rate in the disc.

The picture has points in common with Lynden-Bell's (1979) theory of bar
formation in that it focuses on slowly precessing orbital rings, but it does
not conclude, with Lynden-Bell, that the bar rotates at a rate that
compromises between the fastest and slowest precession rates of the
consituent rings.  Instead we find that both bars and spirals, in common
with  many wave phenomena, move faster than the
underlying medium. Several studies have shown that the rotation rates of
bars in N-body models conform to our predictions rather than those of
Lynden-Bell.

The model of precessing orbits can be introduced directly as is done
in Section 2 after Eq. (\ref{kern}). However, this model is rigorously 
justified by the use of the regular procedure of the perturbation theory 
in the small Lynden-Bell parameter $\epsilon$ from (1), with averaging over 
fast radial oscillations of stars. This programme is implemented in the
beginning of Section 2. The resulting integral equations for disc
normal modes derived in two ways, namely, (i) directly in the model of
precessing orbits, and (ii) with the help of averaging, are identical.

Lynden-Bell (1979) showed that the quantity $J_f = I_1 + I_2/2$ ($I_1$
and $I_2$ are the usual actions) is an adiabatic invariant, provided that
the inequality (\ref{in1}) is satisfied, i.e., for sufficiently slow
perturbations. This means that the orbits evolve by the action of the
potential of low-frequency modes only through change in the angular
momentum $L$, with $J_f$ held constant. 

Assume that in the initial moment $t_0$ the distribution function of
orbits is ${\cal F}_0(J_f,L)$. Let $\Delta L$ be a small variation of
the orbit angular momentum owing to the action of a small potential
perturbation for the time from $t_0$ to the current moment $t$. The
flow of the phase fluid is incompressible, so the phase element with
the initial density ${\cal F}_0(J_f,L-\Delta L)$ should come into the
point $(J_f,L)$ at the moment $t$. Accordingly, the Euler differential
of the distribution function, i.e., its perturbation, is equal to 
\begin{equation}
{\cal F} \equiv {\cal F}_0(J_f,L-\Delta L) -
{\cal F}_0(J_f,L) \simeq -{\cal F}'_0 \Delta L
\label{frch}
\end{equation}
(${\cal F}'_0 \equiv \p {\cal F}_0(J_f,L)/ \p L$).  ${\cal F}$ 
depends on the derivative 
${\cal F}'_0$. So it is not suprising that just this derivative is
found to be critical to our theory of disc slow modes. Henceforth,
this derivative is referred to as the Lynden-Bell 
derivative of the distribution function\footnote{When studying the
  slow bars, Lynden-Bell (1979) has used the same derivative of the
  star precession speed, $\partial \Omega_{pr}/ \partial L$.}.
It turns out that the behaviour of the Lynden-Bell derivative determines
the type of eigen-modes (bar-like or spiral). 

To develop the above
considerations, immediately resulting from the remarkable paper by
Lynden-Bell (1979), into the theory of low-frequency disc modes, one
should invoke the dynamical equations and the expression for the 2D disc
potential (see Sections 2,\,3). As a result, we obtain the integral
equations for these eigen-modes (see Eqs. (\ref{ie1}) and (\ref{ie2})
below). If a potential satisfies these integral equations,
the orbits evolve in such a way that the pattern rotates uniformly,
with a certain angular speed $\Omega_p$.
The physics of our bar-like and spiral solutions consists just in the
fact that these are (slow) normal modes. All specificity (compared
to general modes) is concerned with their slowness. 

The ensuing sections of this paper present the following material.
In Section 2, we derive the basic integral equation of the theory
(by two methods). The basic integral equation is then reduced to the
form of the classical eigen-value problem. Section 3 is devoted to the 
analysis of the general properties of the basic integral equation. In
particular, the crucial role of the Lynden-Bell derivative of the
distribution function is revealed. In Section 4,
this derivative is examined in more detail by the example of the
Schwarzschild model. In two following sections, the specific solutions
of the basic integral equation are given. To demonstrate the
capabilities of the proposed theory, test models studied earlier by
the $N$-body method (Athanassoula \& Sellwood 1986) are adopted
(Section 5). The theory gives results in close agreement with results
of the $N$-body simulations. A wide variety of spiral modes are
studied in Section 6. In the last Section 7, a brief summary of the
results is given. 

\section{Basic equations}
The most convenient variables that should be used under studying the
low-frequency modes are $J_f = I_1 + I_2/2$ 
and $L = I_2$, where $J_f$ is the Lynden-Bell adiabatic invariant
(Lynden-Bell 1979), while $(I_1, I_2) \equiv {\bf I}$ are the usual actions. The
angle variables corresponding to ${\bf I}$ are ${\bf w} \equiv (w_1, w_2)$.
The variables $(J_f, L) \equiv {\bf J}$ are also actions, for which ${\bf\bar
w} \equiv (\bar w_1, \bar w_2) = (w_1,w_2-w_1/2)$ are the canonically
conjugate angles. One can show that for the unperturbed state
\begin{equation}
\bar w_2^{(0)} = \Omega_{pr}({\bf J}) t,
\label{eq2}
\end{equation}
where $t$ is the time, $\Omega_{pr}({\bf J})  = \Omega_2({\bf J}) -
\Omega_1({\bf J})/2$ is the precession speed of the orbit with the actions
${\bf J}$, $\Omega_1$ and $\Omega_2$ are the frequencies of star radial and
azimuthal oscillations: $\Omega_i = \partial H_0({\bf I})/\partial I_i$
$(i=1,2)$, $H_0 = {\bf v}^2/2 + \Phi_0(r)$ is the unperturbed star energy expressed in terms of ${\bf I}$, ${\bf v}$ is the star velocity, $\Phi_0(r)$ is the equilibrium potential. In the limit of near-circular
orbits $\Omega_1 = \kappa(r)$, $\Omega_2 = \Omega(r)$, where $\kappa$
is the epicyclic frequency. 

Eq.\,(\ref{eq2}) means that the angular variable $\bar
w_2$ is slow because $|\Omega_{pr}| \ll |\Omega_1|$, $|\Omega_2|$.
It is easy to show that $\bar w_2 = \alpha$, where $\alpha$ is the azimuth of
the orbit's minor axis (see below, Fig.\,\ref{fig_alpha}).

For later use we give here the coordinates of stars
$r$, $\bar w_2 - \varphi$, expressed in terms of ${\bf J}$,
$\bar w_1$. The radius $r=r({\bf J}, \bar w_1)$ is determined
by solving the equation
\begin{equation}
w_1 (r,{\bf J}) = \Omega _1 \int\limits_{r_{\min}({\bf J})}^r
\frac{\d r'}{\sqrt {2[E({\bf J}) - \Phi_0(r')] -
L^2 /{r'}^2}}
\end{equation}
(it is assumed that $0 \le w_1 \le \pi$).  The slow angular variable
\begin{equation}
\bar w_2 = \varphi + \varphi_1({\bf J}, \bar w_1),
\label{w2e}
\end{equation}
where
\begin{eqnarray}
\varphi_1({\bf J}, w_1) = \Omega_{pr}({\bf J}) \int\limits^r_{r_{\min}({\bf
J})} \frac{\d r'}{\sqrt{2E({\bf J})-2\Phi_0(r') - L^2/{r'}^2}} \nonumber \\
-L\int\limits^r_{r_{\min}({\bf J})} \frac{\d r'}{{r'}^2\sqrt{2E({\bf
J})-2\Phi_0(r') -L^2/{r'}^2}}.
\label{w2b}
\end{eqnarray}

The distribution function $f({\bf J}, \bar {\bf w}, t)$ is governed by the collisionless Boltzmann equation
\begin{equation}
\frac{\p f}{\p t} = [H,f],
\label{cbe}
\end{equation}
where $H$ is the Hamiltonian of a star in the self-consistent
gravitational field $\phi({\bf r},t)$,
$$
H = \frac12 {\bf v}^2 + \phi({\bf r},t),
$$
$[H,f]$ denotes the Poisson bracket,
\begin{equation}
[H,f] \equiv
\frac{\p H}{\p {\bar w}_1} \frac{\p f}{\p J_f} +
 \frac{\p H}{\p {\bar w}_2} \frac{\p f}{\p L} -
\frac{\p f}{\p {\bar w}_1} \frac{\p H}{\p J_f} -
\frac{\p f}{\p {\bar w}_2} \frac{\p H}{\p L}.
\end{equation}

To linearize the collisionless Boltzmann equation, we assume that
$$f = {\cal F}_0({\bf J}) + {\cal F}({\bf J}, \bar w_1) e^{i(m \bar
  w_2-\omega t)},$$
$$\phi = \Phi_0(r) + \Phi({\bf J}, \bar w_1) e^{i(m \bar
  w_2-\omega t)},$$
where ${\cal F}_0({\bf J})=f_0({\bf I})$ and
${\cal F}({\bf J}, \bar w_1)e^{i(m \bar   w_2-\omega t)}$  are the
unperturbed and perturbed distribution functions, respectively, $\Phi({\bf
J}, \bar w_1)e^{i(m \bar
  w_2-\omega t)}$ is the perturbation of the potential, $m$ is the
azimuthal wavenumber, $\omega$ is the frequency ($\omega = m\Omega_p$).
As a result, we obtain the equation
\begin{equation}
-i(\omega - m\Omega_{pr}){\cal F} + \Omega_1\frac{\partial {\cal F}}
{\partial \bar w_1} =
\frac{\partial {\cal F}_0}{\partial J_f}\frac{\partial \Phi}{\partial \bar
w_1} + im\Phi\frac{\partial {\cal F}_0}{\partial L}.
\label{kint}
\end{equation}
To find the desired low-frequency solutions, one can use the
perturbation theory in the small Lynden-Bell parameter $\epsilon$ from (\ref{in1}). 
Let ${\cal F} = {\cal F}^{(1)} + {\cal F}^{(2)}+ \ldots$ be the
perturbation series in powers of $\epsilon$, so that ${\cal
  F}^{(1)}$ is obtained 
from (\ref{kint}) by neglecting the terms proportional to
$\Omega_{pr}$ and $\Phi\propto G$:
$\partial{\cal  F}^{(1)}/\partial {\bar w_1}=0$, i.e. ${\cal
F}^{(1)} = {\cal F}^{(1)}({\bf J})$ is an arbitrary function of the integrals
of motion, which is subsequently specified by using the periodicity condition
for the solution of the next approximation\footnote{Since $|\omega -
m\Omega_{pr}| \ll \Omega_1$, the wave frequency $\omega$ is
not determined in the first approximation.}.

The equation for ${\cal F}^{(2)}$ takes the form
\begin{equation}
-i(\omega - m\Omega_{pr}){\cal F}^{(1)} +
\Omega_1\frac{\partial {\cal F}^{(2)}}{\partial \bar w_1} =
\frac{\partial {\cal F}_0}{\partial J_f}\frac{\partial\Phi}{\partial \bar
w_1} + im\Phi\frac{\partial {\cal F}_0}{\partial L}.
\label{kint2}
\end{equation}
Given the periodicity of functions ${\cal F}^{(2)}$ and $\Phi$,
averaging (\ref{kint2}) over $\bar w_1$ in the interval $(0, 2\pi)$ yields
\begin{equation}
-(\omega - m\Omega_{pr}){\cal F}^{(1)}\approx m\frac{\partial
{\cal F}_0}{\partial L} \bar\Phi \quad(\bar\Phi\equiv\frac1{2\pi}
\int\limits_0^{2\pi}\d\bar w_1 \,\Phi).
\label{kint3}
\end{equation}

The azimuthal number $m$ can take only even values. Formally, it
follows from the relation $\exp(im\bar w_2) = \exp(im w_2)\cdot
\exp(-im w_1/2)$, which remains periodic in $w_1$ (with a period equal to
$2\pi$) only for even $m$.
Physically, a closed precessing orbit is an oval symmetric relative to
the centre. So the torque from the perturbations with odd $m$ will break
rather than rotate such orbits\footnote{It is obvious that the self-consistent perturbations made by such ovals must repeat treir symmetry $ \varphi\to\varphi+\pi $, which is valid only for modes with even $ m $. Among them, the bi-symmetric mode $ m=2 $ is predominant. For instance, just this bar-mode is most likely the only unstable mode. Indeed, the excees of $ \Omega_p $ over $ (\Omega_{pr})_{\max} $ is due to the self-gravitation. This effect is proportional to the function $\Pi$, as it is seen from (\ref{ie2}). From (\ref{pi2}), it follows that the function $\Pi$ decreases with $ m $. So, the growth of the modes with $ m\geq 4 $ owing to CR and OLR would be much less than for the $ m=2 $ mode. The ILR resonance for $ m\geq 4 $, however, will prevent the growth.}.

Calculating the perturbed surface density, $$\Sigma = \int \d {\bf
  v} {\cal F}^{(1)},$$ and using the expression for the 2D disc potential, we
  find
$$
\Phi({\bf r}) = -G\int \d {\bf r}' \frac{\Sigma({\bf r}')}{r_{12}} =
 -G \int \d {\bf r}' \d {\bf v}' \frac{{\cal F}^{(1)}}{r_{12}},
$$
where $r_{12} = [r^2+{r'}^2-2rr'\cos(\varphi' - \varphi)]^{1/2}$. Now
one can change in the last formula from ${\bf r}'$, ${\bf v}'$ to
${\bf J}', {\bar {\bf w}'}$ taking into account that
$\d {\bf r}' \d {\bf v}' = \d {\bf J}' \d \bar {\bf w}'$
($\d {\bf J}' = \d J'_f \d L'$, $\d \bar {\bf w}' = \d \bar w'_1
\d \bar w'_2$):
$$
\Phi({\bf J}, \bar w_1) = -G\int \d {\bf J}'\d \bar{\bf w}'
\frac{{\cal F}^{(1)}({\bf J}') \exp[im\delta\bar w_2]}
{r_{12}},
$$
with $\delta\bar w_2 \equiv \bar w'_2 - \bar w_2$.
Finally, averaging the potential $\Phi$ over $\bar w_1$,we obtain the
following integral equation\footnote{The integration over real $ {\bf J}' $ is correct only for unstable frequencies (Im$\,\omega > 0 $). Thus roots with (Im$\,\omega < 0 $) should be omitted. }:
\begin{equation}
\bar \Phi({\bf J}) = \frac{G}{2\pi}\int \d {\bf J}' \Pi({\bf
J}, {\bf J}') \frac{{\cal F}_0'({\bf J}')}{\Omega_p-
  \Omega_{pr}({\bf J}')} \bar \Phi({\bf J}'),
\label{ie1}
\end{equation}
where ${\cal F}_0'\equiv (\partial {\cal F}_0(J_f,L)/\partial
L)|_{J_f} = \p f_0 /\p I_2 - 1/2 \p f_0/\p I_1$,
\begin{equation}
\Pi({\bf J}, {\bf J}') = \int \d \bar w_1 \d\bar w_1' \d\delta \bar
w_2 \frac{\exp(im\delta \bar w_2)}{r_{12}}.
\label{pi1}
\end{equation}

The function $\Pi$ can be reduced to the following
suitable form:
\begin{equation}
\Pi({\bf J}, {\bf J}') = 8\int\limits_0^{\pi}
\d \bar w_1 \cos m\varphi_1 \int\limits_0^{\pi} \d\bar w_1' \cos m\varphi_1'
\,\psi(r,r'),
\label{pi2}
\end{equation}
where
\begin{equation}
\psi(r,r') = \int\limits_0^{\pi} \d\alpha \frac{\cos m\alpha}
{\sqrt{r^2+{r'}^2-2rr'\cos\alpha}}.
\end{equation}

The function $\Pi({\bf J}, {\bf J}')$ (multiplied by the factor $imG/2\pi$)
has a meaning of torque acting from the harmonical distribution of orbits, $e^{im\bar w_2'}$, with the action ${\bf J}'$ to the orbit with 
the action ${\bf J}$ and orientation $\bar w_2 =0$.

It is notable that the integral equation (\ref{ie1}) in terms of the function
${\cal F}^{(1)}$ has the form of the classical eigen-value problem, where the
eigen-value is directly the pattern speed $\Omega_p$. Indeed, determining 
$\bar\Phi$ through ${\cal F}^{(1)}$
from (\ref{kint3}) and substituting it into (\ref{ie1}), 
one obtains
\begin{equation}
\Omega_p {\cal F}^{(1)}({\bf J}) = \int \d {\bf J}' K({\bf J}, {\bf J}')
{\cal F}^{(1)}({\bf J}'),
\label{ie2}
\end{equation}
with the kernel
\begin{equation}
K({\bf J}, {\bf J}') = \frac{G}{2\pi} {\cal F}_0'({\bf J})
\Pi({\bf J}, {\bf J}') +
\Omega_{pr}({\bf J})\delta[{\bf J}- {\bf J}'].
\label{kern}
\end{equation}
Eq. (\ref{ie2}) is the basic integral equation of the theory.

The integral equation (\ref{ie2}) is much simpler than the general integral
equations for disc normal modes derived earlier by Kalnajs (1965) and
Shu (1970). An ordinary desktop PC is quite capable of solving the integral
equation (\ref{ie2}).

The most important advantage of the present approach is
the fact that it makes clear the underlying physical mechanisms of the
instability processes developing in the disc. To reveal
these physical mechanisms by using the integral equations of Kalnajs or
Shu would be very difficult, and the same is true for the $N$-body
simulations.

The formal derivation of the integral equations
(\ref{ie1},\,\ref{ie2}) for the low-frequency modes of gravitating
discs given above is rigorous. However, the physical situation will be
more transparent after deriving the equivalent integral equation in
another way using explicitly the fact that each orbit as a unit should
take part in slow perturbations of interest.

Accordingly, let us consider the distribution function of closed
precessing orbits, $f(J_f,L;\alpha,t)$, such that $\d{\cal M} =
f\d{J_f}\d L \d\alpha$ is the mass of stars at orbits within a given
interval $\d J_f\d L \d\alpha$, $\alpha$ is the azimuth of minor axis, so that the orbit precession
speed $\Omega_{pr}(J_f,L)=\dot \alpha$.
The collisionless kinetic equation for such a distribution function is
\begin{equation}
\frac{\d f}{\d t} = \frac{\partial f}{\partial t} + \Omega_{pr}
\frac{\partial f}{\partial \alpha} + M\frac{\partial f}{\partial L} =
0,
\label{clke}
\end{equation}
where one takes into account that $(\partial{f}/\partial{J_f})\dot J_f
= 0$ for the slow modes of interest (Lynden-Bell 1979), and $\dot L =
M$, where $M$ is the torque acting on the orbit with given
$J_f,L,\alpha$. Note that such a distribution function $f$ and
kinetic equation (\ref{clke}) were earlier suggested by V.\,Polyachenko (1992).

By linearizing the equation (\ref{clke}) and assuming that $f_1, \bar
\Phi_1 \propto \exp(-i\omega t + im\alpha)$, one can obtain the
equation that is analogous to Eq.(\ref{kint3}):
\begin{equation}
-i(\omega - m\Omega_{pr})f_1 = -M_1\frac{\partial
 {f_0}}{\partial L},
\label{kint7s}
\end{equation}
where $M_1$ is the perturbation of the torque. Let us check
that the equations (\ref{kint7s}) and (\ref{kint3}) are actually
identical. First of all, it is easy to show that
\begin{equation}
M_1 = -\frac{\partial \bar \Phi_1}{\partial\alpha},
\label{mom1}
\end{equation}
where $\bar \Phi_1$ is the potential averaged over the selected orbit;
this potential is produced by all other orbits of a system:
\begin{equation}
\bar \Phi_1 = \bar\Phi_1(J_f, L;\alpha) = \frac1{\mu}\int\limits_0^l
 \d s \, \Phi_1({\bf r})\rho_{lin}^{(J_f, L)}(s),
\label{phibar1}
\end{equation}
where $\rho_{lin}(s)= 1/v(s)$
is the linear mass density characteristic for each
orbit
($v$ is the total star velocity, $s$ is the current length of the
orbit, ${\bf r}$ is the radius-vector at the orbit), $\mu(J_f, L) =
\int\limits_0^l \d s \, \rho_{lin}(s)$ is the mass of stars on the orbit,
\begin{equation}
\Phi_1({\bf r}) = -G\int 
\frac{\d s' \d\alpha' \d J'_f \d L'\rho_{lin}^{(J'_f, L')}(s')f_1(J'_f,
  L';\alpha')}{\sqrt{r^2 + {r'}^2 -
  2rr'\cos(\varphi-\varphi')}} 
\label{phi2}
\end{equation}
($\varphi$ and $\varphi'$ are the current azimuths at two orbits).

In the formulae (\ref{mom1}) and (\ref{phibar1}), we used the same
notation $\bar \Phi$ for the averaged potential as earlier in the
derivation of the integral equation (\ref{ie1}), since in both cases it was
actually the same quantity. To convince oneself that this is correct,
one can immediately compare the two expressions for $\bar \Phi$, taking
into account that a star is within the interval $\d s$ during
$\d t= \d s/v_{tot} = \d r/v_r= \d w_1/\Omega_1$. Thereafter
it remains to make sure that the slow angular variable $\bar w_2$
and the azimuth $\alpha$ of the minor axis are identical.
Indeed, (\ref{w2e}) can be rewritten in more informative manner:
\begin{equation}
\bar w_2 = \varphi -\Delta\varphi + \Omega_{pr}\Delta t,
\label{w2bs}
\end{equation}
where $\Delta t$ is the time it takes for the rotation of a star
through the angle $\Delta \varphi$ between the azimuth of the minor
axis and the current azimuth $\varphi$. Then the identity $\bar w_2 =
\alpha$ is clear from Fig.\,\ref{fig_alpha}. After invoking 
Poisson's equation, we obtain the integral equation that is coincident
with Eq. (\ref{ie1}).

In addition to the derivations above, let us indicate the simplest way to
use Eq.\,(\ref{frch}) of the Introduction for obtaining the relation 
(\ref{kint3}).
The canonical Hamilton equations in variables $({\bf J}, \bar {\bf
  w})$ are
\begin{eqnarray}
\dot J_f &=& -\frac{\partial H}{\partial\bar w_1},\quad \dot I_2 =
-\frac{\partial  H}{\partial \bar w_2}, \nonumber \\
{\dot {\bar w}}_1 &=& \frac{\partial H}{\partial J_f},\quad {\dot {\bar w}}_2 =
\frac{\partial  H}{\partial L},
\label{kaneq2}
\end{eqnarray}
where $H(J_f, L; \bar w_1, \bar w_2)$ is the Hamiltonian.
Since the angular variable $\bar w_2$ is ``slow'', for studying the
low-frequency modes, the procedure of averaging the equations of 
motion over a quick variable (in this case, $\bar w_1$) is
appropriate. As a result, we obtain (see, e.g., Arnold 1989)
\begin{eqnarray}
\dot J_f &\approx& -\frac1{2\pi}\int\limits_0^{2\pi}\d \bar w_1 
\frac{\partial H}{\partial\bar w_1} = 0; \nonumber \\
\dot L &\approx& -\frac1{2\pi}\int\limits_0^{2\pi}\d \bar w_1
\frac{\partial H}{\partial\bar w_2} = - \frac{\partial\bar\Phi_1}{\partial 
\bar w_2}. 
\label{kaneq3} 
\end{eqnarray}
The first equations means the adiabatic invariance of $J_f$, while the
second equation determines the evolution of the angular momentum $L$.
Assuming that $\bar\Phi$ corresponds to a normal
mode, from  (\ref{kaneq3}) one can find:
$$
{\dot L} = \displaystyle - im\bar\Phi\,e^{i(m\bar w_2 - \omega t)} =- im\bar\Phi\,
e^{-i(\omega -m\Omega_{pr})t}.
$$
Integrating the last equation over $t$ from $t_0=-\infty$ to $t$,
we find
\begin{equation}
\Delta L = \frac{m\bar\Phi}{\omega - m\Omega_{pr}} e^{-i(\omega-m\Omega_{pr})t},
\label{delta_l}
\end{equation}
taking into account that the perturbation is switched off at
$t\to -\infty$. Substituting (\ref{delta_l}) into (\ref{frch}) we
obtain (\ref{kint3}). It is clear that the perturbed precession speed is the same 
periodic function of time as $\Delta L$.

\begin{figure}
\begin{center}
\includegraphics[width=7cm]{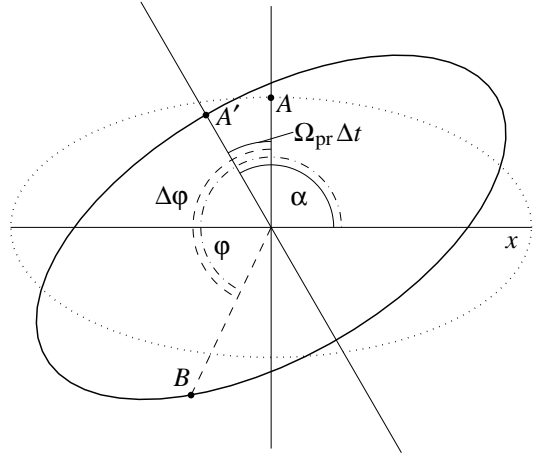}
%\centerline{\psfig{figure=figs/fig7_.eps,width=84mm}}
\end{center}
\caption[]{%
Illustration of the relation (\ref{w2bs}), which proves the identity
$\bar w_2 = \alpha$. Dotted and solid ovals show orientations of an
orbit at times $t=0$ and $t = \Delta t$, respectively; the orbit is
turned by an angle $\Omega_{pr}\Delta t$. The azimuth of a star
initially (at $t=0$) coincides with the azimuth of the minor axis
(point $A$). The position of the star at $t=\Delta t$ is shown by point
$B$; its azumuth is $\varphi = \ang{xB}$ (dash-dotted arc), and the
variation of the azimuth is $\Delta\varphi = \ang{AB}$ (dashed arc).}
\label{fig_alpha}
\end{figure}

Recall that the integral equation (\ref{ie2}) is valid only for even $m$.
However, one can obtain the integral equation describing the low-frequency
modes for $m$ divisible by $m_0$ (where $m_0$ is an arbitrary integer) in the
form (\ref{ie2}), when replacing
$$\bar w_2 = w_2-\frac{w_1}2, \quad
\Omega_{pr} = \Omega_2-\frac{\Omega_1}2, \quad
\frac{\partial {\cal F}_0}{\partial L} = \frac{\partial
f_0}{\partial I_2} - \frac12\frac{\partial f_0}{\partial I_1}
$$
with
$$
\bar w^{(m_0)}_2 = w_2-\frac{w_1}{m_0},
\Omega^{(m_0)}_{pr} = \Omega_2-\frac{\Omega_1}{m_0},
\frac{\partial {\cal F}^{(m_0)}_0}{\partial L} = \frac{\partial f_0}{\partial
I_2} - \frac1{m_0}\frac{\partial f_0}{\partial I_1},
$$
respectively.

The case $m_0=1$ is particularly interesting. The corresponding basic
integral equation is suitable for studying low-frequency modes of discs
around black holes, in planetary rings, and so forth.
Since only two types of the potentials, $\propto r^{-1}$ and
$\propto r^2$ are available, in which all orbits have zero precession speeds,
there are only two cases ($m_0=1$ and $m_0=2$) where our approximation that
$\bar w_1$ is fast is valid in an extensive region.

\section{General analysis of the basic integral equation}
As it follows from (\ref{pi2}), the function $\Pi$ is real and symmetric:
$\Pi({\bf J}, {\bf J}')^*=\Pi({\bf J}, {\bf J}')$,
$\Pi({\bf J}, {\bf J}')=\Pi({\bf J}', {\bf J})$. Let us restrict ourselves to
the regular solutions for the averaged potential $\bar\Phi$. Taking into
account (\ref{kint3}), one can safely divide both sides of (\ref{ie2}) by
${\cal F}_0'$, multiply by ${{\cal F}^{(1)}}^*$, and integrate over
${\bf J}$. Calculating the imaginary part of the resulting equation, one
finds
\begin{equation}
(\mathrm{Im}\,\Omega_p)\cdot \int \d{\bf J} \frac{|{\cal
F}^{(1)}|^2}{{\cal F}_0'}  = 0.
\label{imom1}
\end{equation}
By using (\ref{kint3}), Eq.(\ref{imom1}) can be reduced to the form
\begin{equation}
(\mathrm{Im}\,\Omega_p)\cdot L_m = 0,
\label{imom2}
\end{equation}
where $L_m$ denotes the angular momentum of the mode that can be
obtained from the general expression of Lynden-Bell \& Kalnajs (1972)
retaining the only term which dominates for the low-frequency modes:
\begin{equation}
L_m =  -\int \d{\bf J}
\frac{|{\cal F}^{(1)}|^2}{{\cal F}_0'} = -\int \d{\bf J} {\cal F}_0'
\frac{|\bar\Phi|^2}{|\Omega_p-\Omega_{pr}|^2}.
\end{equation}

It is found that the type of solution of the basic integral equation
depends crucially on the behaviour of the derivative ${\cal F}_0'$.
Below, we consider two cases.

1. Assume that ${\cal F}_0'$ is strictly positive everywhere
in the phase space of the system. Then $L_m$ is negative, as well
as the energy of the mode $\delta E_m$, since $E_m = \Omega_p
L_m$ (Lynden-Bell \& Kalnajs, 1972). Hence, from (\ref{imom2}), one
obtains $\mathrm{Im}\,\Omega_p = 0$. The corresponding real eigen-functions,
${\cal F}^{(1)}$, describe non-spiral solutions. We identify them with the inner parts of bar-modes, i.e., bars themselves.

Obviously, in the case ${\cal F}_0'>0$ under consideration, the integral
equation (\ref{ie2}) determines merely the pattern speed
$\mathrm{Re}\, \Omega_p$ of a mode.
The question now arises of whether there are some external reasons for
growth of bar-modes. It is found that the growth can be due to the exchange by
the angular momentum with the resonance stars at the corotation and
the outer Lindblad resonance (OLR)\footnote{If the disc is immersed
into the real (but not rigid as in the paper by Athanassoula \& Sellwood
(1986)) spherical component, one should take into account, generally
speaking, the resonance angular momentum exchange between the bar-mode and
stars of these component.  The dynamical friction caused by resonance
interactions of stars of spherical systems with a wave was first studied by
Polyachenko \& Shukhman (1982).}. These resonances are located rather
far from the central regions where the modes are mainly localized.
In this respect these resonances are external relative to the modes. Of course,
in the end the correctness of our approach is justified by the comparison 
with results of $N$-body simulations (see Section 5).

The corresponding growth rate can be calculated by the formula
\begin{equation}
\gamma = \frac{\dot L_m}{2L_m},
\label{gr1}
\end{equation}
where $\dot L_m = \dot L^{(1)}_m + \dot
L^{(2)}_m$, and the expressions for the rates of exchange by the
angular momentum at CR ($\dot L^{(1)}_m$) and OLR ($\dot
L^{(2)}_m$) can be obtained from the general formulas of
Lynden-Bell \& Kalnajs (1972) through their minor transformation:
\begin{equation}
\dot L^{(l)}_m =
\frac1{4\pi}\int \d{\bf J}
\left(\frac{l}2\frac{\partial {\cal F}_0}{\partial J_f} +
\frac{\partial {\cal F}_0}{\partial L}\right) |\Phi^{(l)}_1|^2
\delta[\Omega^{(l)}({\bf J}) - \Omega_p],
\label{er1}
\end{equation}
where $\delta$ is the Dirac delta-function,
$\Omega^{(l)}({\bf J}) \equiv \Omega_2({\bf J}) +
 (l-1)\Omega_1({\bf J})/2$, 
and
\begin{equation}
\Phi^{(l)}_1 =
-\frac{G}{\pi} \int  \d{\bf J}' \d\bar w_1 \d\bar w_1'
\psi(r,r') {\cal F}^{(1)}({\bf J}') e^{-il\bar w_1 +
2i(\varphi_1'-\varphi_1)}
\end{equation}
is the Fourier coefficient corresponding to CR ($l=1$) and OLR
($l=2$) in the expansion of the potential in the series in $e^{il\bar
  w_1}$.

The interaction of stars in the resonance regions
with the gravitational potential of the mode leads
to the spiral responses. For the corotation resonance, as an example,
the response consists of two parts. The resonance
response of the disc, without taking its self-gravity into account,
produces spirals of angular length about $\pi/2$ (see, e.g., Polyachenko
2002a). The self-gravity also extends the resonance spirals by
$\pi/2$ (Polyachenko 2002b). Therefore, the total angular length is about
$\pi$, which is typical for the majority of SB galaxies (see, e.g.,
Sandage 1961).

It should be emphasized that above we consider only the linear stage
of the formation of structures in SB galaxies. The observed
galaxies can have strongly nonlinear bars. The length of such bars is
apparently determined by 4:1 resonance. This is exactly the case,
e.g., for the nonlinear bar-modes computed by Sellwood \& Athanassoula
(1987). For a symmetric bar which is free of odd Fourier harmonics, in
particular $m=3$ harmonic, the resonance 4:1 is the nearest to the
centre, and thus is the most important one.

2. Let us say that ${\cal F}_0'$ becomes negative in some regions of the phase
space.  For the realistic distribution function (see Section 4), these
regions, if they exist, can occupy only a small fraction of the total volume
of the phase space.  Then, in addition to the fast bars described above, the
spiral-like solutions can occur. Contrary to the bars, these new modes grow
due to their intrinsic instability, which is inherent to the mode itself.  As
follows from (\ref{imom2}), for an unstable mode ($\mathrm{Im}\,\Omega_p
> 0$), the angular momentum $L_m=0$, i.e. the contributions to $L_m$ from the
regions with opposite signs of the Lynden-Bell derivative ${\cal F}_0'$
cancel each other exactly: $L_m = L^+ + L^- = 0$.  Obviously, the instability
is due to the inner Lindblad resonance, since the criterion of 
instability, ${\cal F}'_0<0$, coincides exactly  with the Lynden-Bell \& Kalnajs condition of wave exitation at ILR (see (\ref{er1}) when $l=0$). This resonance becomes the source of
spiral waves in the case under consideration, contrary to the commonly
accepted view of the role of ILR.  Accordingly, the growth rate of
the unstable mode can be calculated as
\begin{equation}
\gamma = \frac{\dot
L^+}{2L^+} = \frac{\dot L^-}{2L^-},
\label{gamma2}
\end{equation}
where
\begin{equation}
\dot L_\pm = \frac1{4\pi} \int\limits_{\Gamma_\pm}\d{\bf J}
{\cal F}_0'|\bar\Phi|^2\delta[\Omega_{pr}({\bf J})-\Omega_p],
\label{lpm}
\end{equation}
$\Gamma_+$  and $\Gamma_-$  denote the regions of the phase  space with
positive and negative values of the
Lynden-Bell derivative ${\cal F}_0'$, respectively.
Note that below the mode growth rates are obtained
directly from the solution of the integral equation (\ref{ie2}) as
$\gamma = 2\mathrm{Im}\, \Omega_p$. However, it is also useful
to derive a simple formula for estimate of the growth rate from
(\ref{gamma2}, \ref{lpm}).
The simplest case is that with one narrow domain of the negative derivative
${\cal F}_0'$. Using the latter equality of (\ref{gamma2}) and approximating
the integrals as the products of the average values of the integrands and
small volume of the phase space domain, we obtain:
\begin{equation}
\gamma \simeq |\Omega_{pr}'|\Delta L,
\end{equation}
where $\Delta L$ is the width of the region with ${\cal F}_0'<0$, and
$\Omega_{pr}'\equiv \partial\Omega_{pr}/\partial L$ is calculated at the
narrow region of the phase space of interest.
This estimate is consistent with the explanation of the physical
mechanism of the angular momentum exchange at the ILR given by Lynden-Bell \&
Kalnajs (1972).
However, two more points need to be made: (i) the sign of the effect
is opposite to that in the cited paper,
in accordance with the fact that ${\cal F}_0'<0$, and
(ii) strictly speaking, their explanation is appropriate only for the case
of near-circular star orbits, although, as we demonstrate below, the very
elongated orbits play an essential role. Nevertheless, their considerations
can readily be generalized to such a case.

\section{Examination of the Lynden-Bell derivative by the example of the
typical model}
Let us consider the generalized Schwarzschild distribution function
(Shu 1970)
\begin{equation}
f_0(E,r_0) = \frac{2\Omega(r_0)}{\kappa(r_0)}
\frac{\sigma_0(r_0)}{2\pi c_0^2(r_0)}
\exp\left(-\frac{E-E_c(r_0)}{c_0^2(r_0)} \right),
\label{schw}
\end{equation}
where $E$ is the star energy, $r_0$ is the radius of the guiding centre:
$L=r_0^2\Omega(r_0)$,  $E_c(r_0)=v_0^2(r_0)/2+\Phi_0(r_0)$ is the star energy
at the circular orbit, $v_0(r_0) = r_0\Omega(r_0)$ is the circular velocity,
$\Phi_0(r_0)$ is the equilibrium potential. The specific model is
given by the functions $\sigma_0(r_0)$ and $c_0(r_0)$; in the epicyclic
limit, when $v_0/c_0 \gg 1$, $\sigma_0(r_0) = \Sigma_0(r_0)$, $c_0(r_0) =
c_r(r_0)$, where $\Sigma_0(r_0)$ and $c_r(r_0)$ are the surface density and
radial velocity dispersion respectively. In the general case,
$\Sigma_0(r_0)$ and $c_r(r_0)$ are expressed by $\sigma_0(r_0)$ and
$c_0(r_0)$ in a more complicated manner.

The Lynden-Bell derivative of the distribution function (\ref{schw}) is equal
to
\begin{eqnarray}
\frac{\partial {\cal F}_0^{(m_0)}}{\partial L}
 = \frac{2\Omega(r_0)}{r_0^2\kappa^2(r_0)}{\cal F}_0
\left\{r_0\frac{\Omega'(r_0)}{\Omega(r_0)} -
r_0\frac{\kappa'(r_0)}{\kappa(r_0)} +
r_0\frac{\sigma_0'(r_0)}{\sigma_0(r_0)} - \right. \nonumber \\
\left.
r_0\frac{2c_0'(r_0)}{c_0(r_0)} +
\frac{r_0^2\kappa^3}{2m_0\Omega(r_0)c_0^2(r_0)}
+ r_0\frac{2c_0'(r_0)}{c_0^3(r_0)}(E-E_c(r_0)) \right\},
\label{lbd1}
\end{eqnarray}
where ${\cal F}_0({\bf J}) = f_0(E({\bf J}),r_0(L))$,
a prime denotes the derivative with respect to $r_0$. As a rule,
${\cal F}_0'>0$ either in all the phase space or at least for the most part
(almost everywhere). It is provided by the term
$r_0^3\kappa^3/2m_0\Omega c_0^2$ in (\ref{lbd1}). For example, in the case
of the flat rotation curve ($v_0=\mathrm{const}$), this term equals 
$(\sqrt2/m_0) (v_0/c_0)^2$. Since $v_0/c_0 \gg 1$  almost
everywhere in disc galaxies, this term is dominant.

However, there may exist some narrow regions, in which such predominance
is broken. Firstly, this is possible in the regions where the function
$\kappa(r)$ is sufficiently small. For instance, with a rotation curve
similar to that of our Galaxy, such regions can be located at $r
\approx 2.5$ kpc and $8-10$ kpc. Note the obvious causes of such
peculiarities in the rotation curves: (i) a changeover from the
potential of the spherical component to the potential of the disc, and
(ii) a sharp edge of one of the disc components.  Secondly, the
function $c_0(r_0)$ increases rapidly when approaching the galactic
centre, so that the rotation velocity $v_0$ in the central parts
becomes of the order of $c_0$. 

Relatively small deviations in the behaviour of the rotation curve may lead
to very different solutions of the basic equation.
Some other factors are also very important. The excess of the pattern speed above the maximum precession rate is due to self-gravity. Thus the solutions have the bar-like form only in sufficiently massive discs, while for discs with relatively small masses we can obtain only spiral modes. The number of
very elongated orbits plays the essential role. For example, the excess of
such orbits over the number prescribed by the generalized
Schwarzschild distribution function (\ref{schw}) strongly increases the
growth rate of the spiral modes due to increase of
$|\Omega_{pr}'|$ for these orbits. Besides, it is seen from (\ref{lbd1}) that
$|{\cal F}_0'|$ (${\cal F}_0'<0$) increases with $(E-E_c(r_0))$.

\section{Numerical test solutions of the basic integral equation}
The basic integral equation is studied numerically. The unknown function
${\cal F}^{(1)}({\bf J})$ as well as the kernel $K({\bf J}, {\bf J}')$ are
considered on the $31\times 31$ grid in the phase space $(E,L)$. Then the
obtained matrix equation can be solved using standard methods of linear
algebra.

It is advisable to put aside the distribution function (\ref{schw}) until
the next section and demonstrate the capabilities of the theory by
applying it to test models. For such a role, we choose two models investigated by
the $N$-body simulations of Athanassoula \& Sellwood (1986).
The equilibrium potential for all the models is the Plummer potential,
$\Phi_0(r) = -(1+r^2)^{-1/2}$. 

\begin{figure}
\begin{center}
\includegraphics[width=8.4cm]{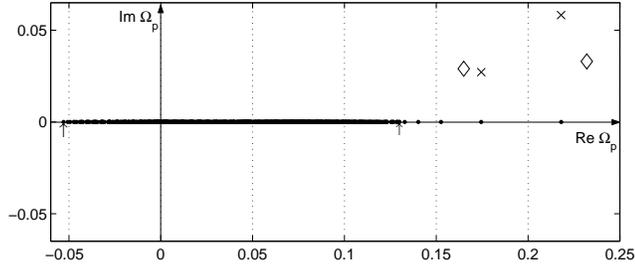}
%\centerline{\psfig{figure=figs/fig1.eps,width=84mm}}
\end{center}
\caption[]{%
The spectrum of pattern speeds (filled circles) computed as the
eigen-values of the problem (\ref{ie2}), for one of the Kalnajs models. The
crossess are the complex pattern speeds, with the growth rates estimated by
the formulas (\ref{gr1}), (\ref{er1}). The diamonds are the ``experimental"
complex pattern speeds according to Athanassoula \& Sellwood (1986).
The arrows point to the minimum and maximum values of the star precession
speed.}
\label{fig1}
\end{figure}

First of all, we are dealing with a Kalnajs (1976) model, which
Athanassoula \& Sellwood (1986) have denoted by ($m=6$, $\beta =0$,
$q=1$, $J_c=0.25$) (see their Table 1). 
The computed spectrum of the matrix equation, which consists of $31^2$ 
eigen-values, is shown in Fig.\,\ref{fig1}. The frequencies are
real, since the Lynden-Bell derivative for the Kalnajs models is positive
everywhere in the phase space. Just few of eigen-values correspond to the 
discrete spectrum, while 
others mimic the continuous part of the spectrum\footnote{The existence 
of the continuous part of the spectrum is a property of the 
kernel (\ref{kern}), which defines a bounded noncompact integral 
operator ({\it cf.} Reed \& Simon 1972).}, which is the van Kampen -- Case 
modes (van Kampen 1955; Case 1960).  They are located in the interval 
between the minimum and maximum values of the precession speed: 
$(-0.051, 0.126)$. These modes are the stable solutions 
of the form $\propto \delta({\bf J} - {\bf J}_0)$. The peculiar
van Kampen modes are of no interest in studying the bar modes.
Thus, in the
$N$-body simulations of Athanassoula \& Sellwood, the bar is always one of
the modes of discrete spectra. The latter are the eigen solutions of the
basic integral equation. Both the perturbed potential and surface density of
the discrete modes differ in the number of radial nodes, the nodeless mode
being the mode with the maximum pattern speed.

In the spectrum given in Fig.\,\ref{fig1} one can see 5 discrete
modes. The figure shows that the pattern speeds of the first and
second modes (counting from the right) coincide with the pattern speeds
obtained in the $N$-body simulations to within 6\%. It is found (see
below) that the growth rates of the other 3 modes are much lower, so
they could not be observed in the numerical experiments.

Rough estimates of growth rates for the most rapid modes, in which we
assumed that the star orbits are circular, give $\gamma_1 = 0.117$,
$\gamma_2 = 0.054$.
Each of these values is composed of two parts corresponding to
corotation (CR) and OLR: $\gamma_{1CR} = 0.0199$, $\gamma_{1OLR} = 0.0968$,
$\gamma_{2CR} = 0.0107$, $\gamma_{2OLR} = 0.0437$. Note that in this
case, the OLR terms are dominating.
The mode with the maximum pattern speed has the highest growth
rate. Obviously, the reason is that such a mode possesses the smallest
corotation radius, where the disc surface density is high.

The estimate for the growth rate of the most rapid mode is nearly
twise as much as one obtained by Athanassoula \& Sellwood (1986) in
the $N$-body simulations. For the second mode, the ``experimental'' and
``theoretical'' growth rates are quite close. Among the reasons of the
possible discrepancy, one can mention the use of the circular orbit
approximation. Recall that the estimates obtained by Athanassoula \&
Sellwood (1986) within the swing amplification approach are twice
as little than their ``experimantal'' values.

\begin{figure}
\begin{center}
\includegraphics[width=8.4cm]{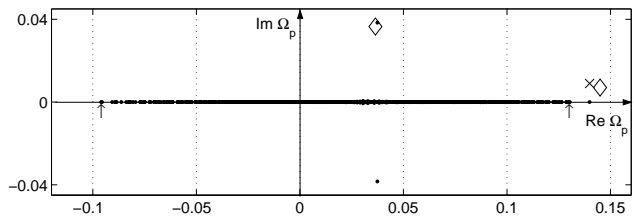}
\end{center}
\caption[]{%
The spectrum of complex pattern speeds (filled circles) computed as the
eigen-values of the problem (\ref{ie2}), for the model ($m=6$, $\beta
=3$, $q=1$, $J_c=0.6$) of Athanassoula \& Sellwood (1986). The
cross is the complex pattern speed, with the growth rate estimated by
(\ref{er1}). The diamonds are the ``experimental" complex pattern
speeds according to Athanassoula \& Sellwood. The arrows point to the
minimum and maximum values of the star precession speed.}
\label{fig5}
\end{figure}

Another model was denoted by Athanassoula \& Sellwood as ($m=6$,
$\beta =3$, $q=1$, $J_c=0.6$)\footnote{The pattern speeds for
other models from the list of Athanassoula \& Sellwood (1986) were also
computed as the eigen-values of the problem (\ref{ie2}) (about ten models in
all). In all cases the results coincide very closely (with an accuracy of
10\% or better). Besides, the eigen-frequencies of the basic integral
equation for the isochrone potential coincide within a few percents
with those obtained by Kalnajs (1978) using his matrix equation.}.
The computed spectrum of eigen-values of bi-symmetric modes for this model,
is shown in Fig.\,\ref{fig5}.
The continuous spectrum of van Kampen modes is again located in the
interval between the minimum and maximum values of the precession speed:
$(-0.096, 0.13)$.

\begin{figure}
\begin{center}
\includegraphics[width=7cm]{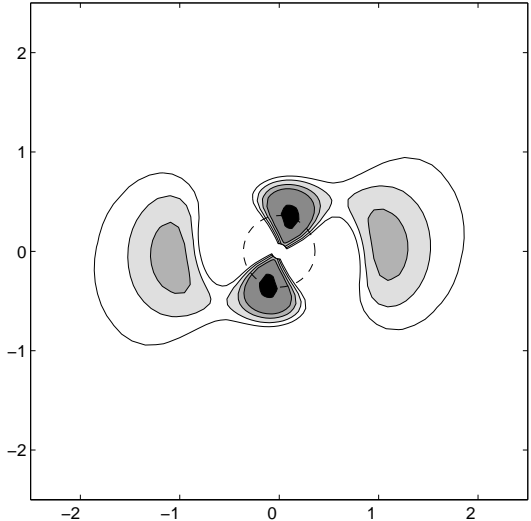}
\end{center}
\caption[]{The spiral pattern of the unstable mode in the model
  ($m=6$, $\beta =3$, $q=1$, $J_c=0.6$) of Athanassoula \& Sellwood
  (1986). The dashed circle indicates
  the position of the inner ILR (the outer ILR, for this model, is at
  $r=5$).}
\label{fig6}
\end{figure}

Since the Lynden-Bell derivative for the model under consideration is
not positive
everywhere in the phase space, the discrete spectrum contains
complex modes. In Fig.\,\ref{fig5}, we can see one unstable complex
and one real discrete frequencies. The complex
eigen-frequency obtained from the solution of the basic integral
equation is very close to the ``experimental'' frequency, given by
Athanassoula \& Sellwood (1986). The spiral pattern of the unstable mode
is given in Fig.\,\ref{fig6}.

The pattern speed of the real mode coincides with the pattern speed
obtained in the $N$-body simulations to within 3\%. The estimate of
its growth rate, in the circular orbit approximation, gives $\gamma =
0.018$, the OLR contribution being dominating. This estimate is 30\% larger than
that obtained by Athanassoula \& Sellwood (1986) in the $N$-body
simulations.

The models explored above are rather artificial. Nevertheless, they were
convenient for us
because it was thoroughly studied in the $N$-body simulations by
Athanassoula \& Sellwood (1986). Besides, the analysis above points to
the mechanism of bar-formation in the general case. It is clear that bars can
form in galaxies with sufficiently massive discs. In this case, the necessary
inequality $\Omega_p > (\Omega_{pr})_{\max}$ can be fulfilled, otherwise the
inner Lindblad resonances occur. Then, in the case of a positive Lynden-Bell
derivative, only the continuous spectrum of the van Kampen
waves exists. Note that the fulfillment of the inequality
$\Omega_p > (\Omega_{pr})_{\max}$ is made difficult not only for low-mass
discs but also for discs with a high concentration of mass at the centre.

The permitted frequency (pattern speed) of the galactic bar must be one
of the eigen-values of the basic integral equation. It means that the
corresponding waves remain unchanged over many revolutions of the galaxy,
despite the different rates of orbit precession. 
The half-turn spirals adjacent to the bar form due to the same resonance
interaction.

\section{Spiral solutions of the basic integral equation}
Here we consider in greater detail the situations that arise when the
Lynden-Bell derivative
is negative in some regions of the phase space. Then a great number of
available possibilities appears; they are briefly outlined below.
An extended discussion will be published elsewhere.

\begin{figure}
\begin{center}
\includegraphics[width=8.4cm]{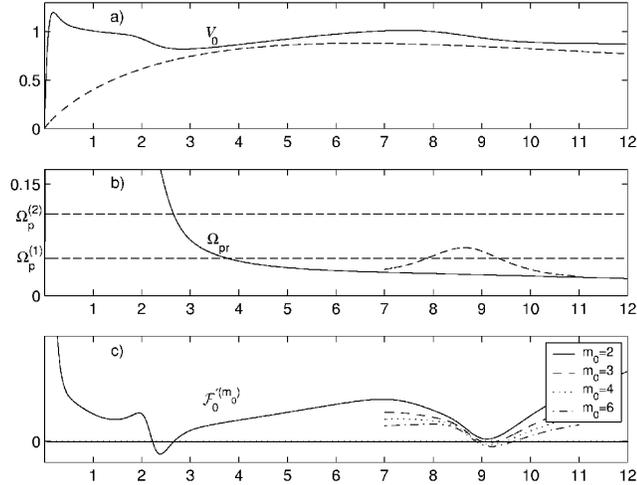}
%\centerline{\psfig{figure=figs/fig2.eps,width=84mm}}
\end{center}
\caption[]{(a) The ``Milky Way-type" rotation curve (solid line); dashed line
corresponds to the contribution of the exponential disc; (b) The precession
speeds, $\Omega_{pr}$, of the circular orbits for the standard model (solid
line) and the model with a ``peak" near $r=8-10$ (dashed line). The
horizontal straight lines correspond to different pattern speeds; (c) The
Lynden-Bell derivatives, ${{\cal F}_0'}^{(m_0)}$, for the
circular orbits.}
\label{fig2}
\end{figure}

As a base distribution function for the analysis of the spiral solutions of
the basic integral equation, we choose the generalized Schwarzschild
distribution function described in Section 4. For the standard model, we
assume the ``Milky Way-type" rotation curve (see Fig.\,\ref{fig2}a), and the
exponents $\sigma_0(r_0) = \sigma_0 e^{-r_0/r_d}$,
$c_0(r_0)=c_0 e^{-r_0/r_c}$ that fix the particular Schwarzschild
function. In our units, one kpc is equal to 1, $r_d = 3$,
$r_c=2r_d$, the gravitational constant $G = 1$, $\sigma_0 = (\pi r_d)^{-1}$.
The magnitude of the rotation curve is chosen to match the disc component
of the velocity curve at radii $r_0 \gtrsim 5$. The maximum rotation velocity is
$(v_0)_{\max}=1.2$, $c_0 \simeq 0.83$, so that for $r_0=8$ the ratio
$v_0(r_0)/c_0(r_0)\simeq 4$. For such parameters, the galactic disc is
exponential almost everywhere, excluding the very central part. Given the
rotation curve, the equilibrium potential is $\Phi_0(r) = \int\limits^r
\d r' \, v_0^2(r')/r'$. In such a manner we obtain the model of a galaxy, which
is similar to our Galaxy (but, of course, the model galaxy is not 
identical).\footnote{Remind that the occurrence of the grand design spiral structure in the Galaxy is so far open to question. The galaxies, which we mean in our theory, are more in the nature of NGC~2997 (see, e.g., the image of this galaxy on the cover of the book by Binney \& Tremaine (1987).}

\begin{figure}
\begin{center}
\includegraphics[width=8.4cm]{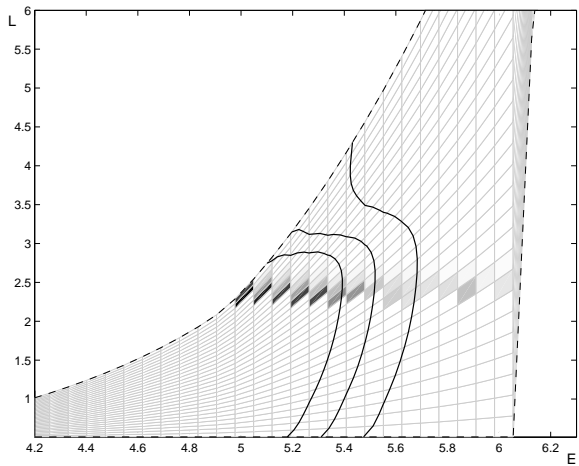}
%\centerline{\psfig{figure=figs/fig3.eps,width=84mm}}
\end{center}
\caption[]{The strip of negative values of the Lynden-Bell derivative,
${\cal F}_0'$, at the phase plane $(E,L)$. The phase domain is bounded by
the line of circular orbits (from the top) and the line corresponding to
$0.02\cdot({\cal F}_0)_{\max}$, where $({\cal F}_0)_{\max}$ is the maximum
value of the distribution function, ${\cal F}_0(E,L)$, at each fixed value of
$E$ (from the right). The solid curves, $\Omega_{pr}(E,L)=\Omega_p$, indicate the
locations of resonance orbits on the phase plane, for several values of
$\Omega_p$.}
\label{fig3}
\end{figure}

Fig.\,\ref{fig2}c shows the Lynden-Bell derivative, ${\cal F}_0'$, for
the standard model calculated by (\ref{lbd1}) at the line of circular orbits
($E=E_{cr}(r_0)$). There is only one narrow region with ${\cal F}_0' < 0$
located near the centre. Fig.\,\ref{fig3} demonstrates how
deep into the phase space the negative values of ${\cal F}_0'$ penetrate.

The basic integral equation for the standard model provides a large number of
unstable modes. They can be either trailing or leading spirals, the latter
are the majority. Narrowly-localized van Kampen solutions of the continuous
spectrum, which occupy the wide band of real frequencies, $\Omega_{pr}^{\min}
< \Omega_p < \Omega_{pr}^{\max}$, can be both the trailing and leading spirals
since for each of these modes, effectively, the gravitational constant $G\to
0$\footnote{As computations show, the van Kampen modes are the sole type of
solutions for the case when ${\cal F}_0' >0$ everywhere in the galactic
disc with the assumed parameters.}. When employing the concept of
eigen-modes, just the packets of leading van Kampen waves participate in
the classical swing amplification process described in the original papers by
Goldreich \& Lynden-Bell (1965), Julian \& Toomre (1966), and Toomre (1981).
If the dispersion of orbit speeds, $\Delta\Omega_{pr}$, is sufficiently
small (that is always assumed), we obtain quasi-monochromatic packets of
van Kampen waves. Just such wave packets moving with some group velocity
from the central regions and approaching the corotation, can be
amplified by the resonance effects under the transformation from the leading
form to trailing one. Though each of the van Kampen modes is essentially
kinematic ($G\to 0$), the finite packets of these waves are governed by their
self-gravity. In particular, the local dispersion relation of
Lin--Shu--Kalnajs should be valid for them (away from the resonances).

Here we touch on those improvements,
which the present theory can introduce into the original scheme of the swing
amplifier. They are connected with the possibility of unstable discrete
leading spiral modes. In particular, there is then no need to reflect the
short trailing waves somewhere in the central regions to close the
feedback loop. Given that the discrete leading modes are monochromatic,
the efficiency of the resonance amplification should grow in comparison with
quasi-monochromatic packets of the van Kampen waves. These modes might be the
``embryo" waves for their subsequent resonance growth due to the swing
amplifier. Then, apparently, from the full set of these waves, the mode with
the maximum potential perturbation at the corotation is selected. Let us
emphasize again the ``inversed" role of the ILR (compared to the commonly
accepted one), which acts as the generator of spiral waves.

The growth rates of spiral modes for the standard model are rather low:
they provide the growth of the amplitude $e$ times, typically
for $(3-5)\cdot10^9$ years (taking into account
that $(v_0)_{\max} = 1.2$ corresponds to
the rotation period $T=2.5\cdot10^8$ years). The main reason for such low
values of growth rates follows from Fig.\,\ref{fig3}. This figure shows
that the curve $\Omega_{pr}(E,L) = \mathrm{Re}\,\Omega_p$ (for a typical
unstable spiral mode) intersects the strip of negative values of the
Lynden-Bell derivative, ${\cal F}_0'$, deep inside the phase space of the
system under consideration, at $(E,L)$ corresponding to highly elongated
orbits. But
the number of such orbits in the Schwarzschild distribution function
(decreasing exponentially with the ratio $(E-E_c(r_0))/c_0^2(r_0)$) is small.
The growth rates of the spiral modes can be significantly increased if one
takes into account that the number of such orbits is actually very much
larger.  Indeed, we are dealing with the central regions of spiral galaxies,
where the contribution of the spherical component, i.e. stars of Population
II that have the radially elongated velocity diagram, is
essential. It implies that the spherical component can play an
important active role, instead of the passive role of a rigid halo
as is commonly
accepted. Computation with the standard model that is accordingly
modified, leads to tenfold growth rates. The modification consists in the
substitution $c_0(r_0) \to 2c_0(r_0)$ provided that $(E-E_c(r_0)) >
c_0^2(r_0)$. For such a modification, the observed surface density and radial
velocity dispersion are almost unchanged. It is also important that this
modification leads to the predominance of trailing spiral modes. The observed
galactic spirals are most likely just these modes (or similar spirals in
other modifications of the standard model). For instance, the modes are most
often the trailing half-turn spirals as in the majority of observed galaxies.

\begin{figure}
\begin{center}
\includegraphics[width=8.4cm]{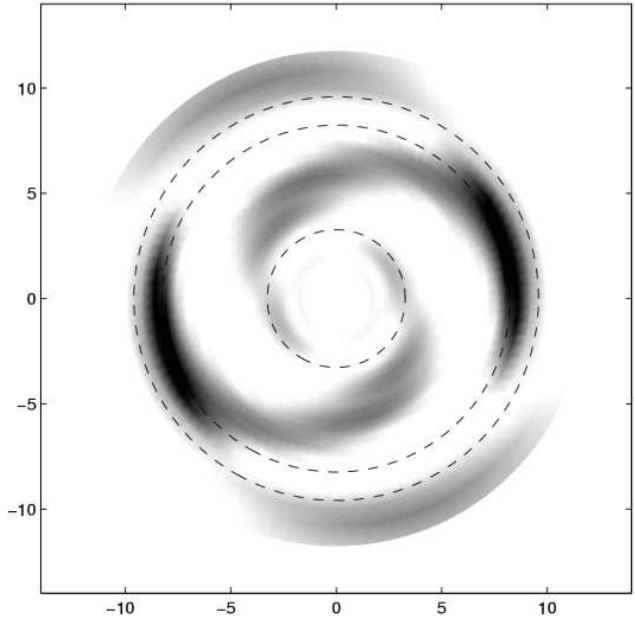}
%\centerline{\psfig{figure=figs/fig4.eps,width=84mm}}
\end{center}
\caption[]{One of typical pictures of unstable spiral modes.}
\label{fig4}
\end{figure}

In particular, an interesting modification consists of forming the
``peak"\footnote{For our Galaxy, such a ``peak" has been
marked since the well-known
paper by Lin, Yuan, \& Shu (1969).} on the curve $\Omega_{pr}$ at sufficiently
large $r$, where the Lynden-Bell derivative can be either positive or
negative (in Fig.\,\ref{fig3}, the upper right corner of the $(E,L)$
phase plane). We omit descriptions of the variety of cases, restricting
ourselves to the only picture of the typical spiral mode (Fig.\,\ref{fig4}).

In conclusion let us touch on one very important aspect of the problem.
Fig.\,\ref{fig2}c shows that the ``generalized''  Lynden-Bell derivative
${{\cal F}_0'}^{(m_0)}$ at $m_0>2$ can be made negative more easily
compared to the case $m_0 =2$. This effect is due to
the inverse dependence of the large positive term
$r_0^2\kappa^3/2m_0\Omega c_0^2$ in the expression (\ref{lbd1})
on $m_0$. Hence, in the outer parts of galaxies (for our Galaxy, just
beyond the Solar circle) the additional multi-arm structures are generated,
which are independent from the central spiral structure (the latter usually
has two
arms)\footnote{The curves ${{\cal F}'_0}^{(m_0)}(r)$ in Fig.\,\ref{fig2}c are
shown only at sufficiently large radii to emphasize that modes with $m_0>2$
cannot typically occupy as wide a region as the two-arm mode does. This is
explained by the different behaviour of $\Omega_{pr}(r)$ and
$\Omega_{pr}^{(m_0)}(r)$ for $m_0>2$.}.
Furthermore, one can expect that
several modes will be generated there simultaneously leading to formation of
the so-called flocculent structures (Elmegreen \& Elmegreen 1982, 1987).  All
the aforementioned agrees with observations of galaxies (see, for example, The
Hubble Atlas of Galaxies, Sandage (1961)).  As a rule, in the central part of
the disc, the regular two-arm spiral is seen, which is then
superseded by a multi-arm or disordered  structure. It is appropriate to
mention here that Elmegreen (1992) has clearly demonstrated that three-arm
spiral structures appear only on the periphery of galaxies beyond the inner
Lindblad resonance for the $m=3$ mode.

\section{Discussion}
The new approach to the study of structures in spiral galaxies suggested above
is based on a view of a galactic disc as a set of precessing orbits.
Following this model, we have obtained the simple integral equation that is
a very convenient tool
for studying galactic structure. Non-axisymmetric structures are considered as
the normal modes of the basic integral equation (\ref{ie2}). It is found
that the properties of solutions depend crucially on the behaviour of the
Lynden-Bell derivative, ${\cal F}_0'$, at the phase space of the disc. The
solutions have bar-like or spiral form depending on whether ${\cal F}_0'$
is positive everywhere or negative somewhere in the phase space.

Within the framework of the suggested
approach, which does not include the swing amplifier, we succeeded in
calculating the pattern speeds and growth rates of all the modes obtained
earlier in the $N$-body simulations by Athanassoula \& Sellwood (1986), the
coincidence of results being very high.

Thus, it is shown that the bar-mode grows due to the immediate action of its
gravity on the stars in the vicinity of corotation leading to the angular
momentum exchange between these stars and the bar-mode.
Note that Athanassoula \& Sellwood (1986) could readily calculate
exact growth rates of all the modes they computed by the formulas
(\ref{gr1}, \ref{er1}). However, they preferred the language of swing
amplification for rough estimates of the growth rates.

To avoid confusion, we should emphasize that here we deal only with the
standard fast bars. As for the slow bars of Lynden-Bell (1979), they are most
likely the central parts of the unstable spiral modes (in this sense, they
are secondary). Note that here we leave aside the possibility of the slow bar
formation in very hot centres of galactic discs (in full analogy to the
ellipsoidal deformation of spherical systems under the action of the radial
orbit instability). In contrast, the fast bars are primary while the
adjacent spirals are secondary since the spirals form as the response of the
disc near the corotation to the bar gravity.

The commonly used approach to the problem of
formation of the normal spiral structures consists in using either the
swing amplification mechanism (Toomre 1981) or its weaker counterpart
(overreflection with reference to the waser mechanism -- see, e.g.,
Bertin \& Lin 1996).
Note that the swing amplification mechanism is sometimes the only way for
growth of the initial perturbations. It seems likely that such galaxies
do not have any organized spiral structure of the mode nature.
Certainly, this mechanism can effectively amplify
transient wave perturbations.

In traditional mechanisms, the region of corotation plays the central
role. In our approach, we draw  attention to ILRs that can under certain
conditions cause the excitation of a great variety of spiral modes. Unlike
the usual epicyclic idea of ILR as a certain circle, ILR in our unstable
modes extends over wide regions of the galactic disc. This is due to
participation of a variety of orbits including highly elongated orbits (see
Fig.\,\ref{fig3}).  On the one hand, ILR can give rise to leading spiral
modes, which in turn can be the ``embryos" for further work of the swing
amplifier.  On the other hand, ILR can provide a variety of trailing spirals
including those in the region between two ILRs (see Fig.\,\ref{fig2}b).

\section*{Acknowledgments}

With thanks to V.L. Polyachenko for useful discussions. 

\section*{Appendix. Some additional arguments in support \\ of the used approach}

1. The perturbed distribution function $ f $ is represented as a Fourier series that contains resonance denominators: 
\begin{equation}
f \sim \sum\limits_l \frac{a_l\,e^{ilw_1}}{(\omega - m\Omega_2) + l\Omega_1}.
\label{dfe1}
\end{equation} 
It is natural to name the terms with $ l = -1 $,  $l=0$, and $ l=1 $ as ``ILR'', ``CR'', and ``OLR'' terms, respectively. In Section 2, we have restricted ourselves to the only ILR term. As it was shown in Section 4, this leads to the correct results (with an accuracy of 10\%) in determination of the pattern speeds. Thus, one can conclude that the ILR-term has a dominant role.

\begin{figure}
\begin{center}
\includegraphics[width=8cm]{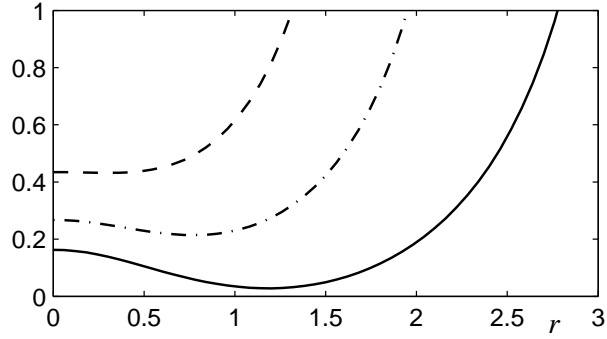}
\end{center}
\caption[]{%
The ratio $A(r)$ for different values of the pattern speed $ \Omega_p $: solid,  dash-dotted, and dotted lines correspond to the minimum ($ \Omega_p^{\min}=0.14 $), mean ($ \bar\Omega_p=0.21 $), and maximum ($ \Omega_p^{\max} = 0.3 $) pattern speeds, respectively.} 
\label{fig1s}
\end{figure}

\vspace{12pt}\noindent
2. The ratio of the denominator of the ILR term to the denominator of the CR term (closest to ILR) is
\begin{equation}
A = \frac{|\omega-m\Omega_{pr}|}{|\omega - m\Omega_2|} = 
\frac{\delta\Omega}{|\Omega - \Omega_p|},
\label{ratio}
\end{equation} 
where the quantity $ \delta\Omega = |\omega-m\Omega_{pr}| $ was already used in the Introduction. Strictly speaking, this ratio is equal to the Lynden-Bell small parameter $ \epsilon $ from (\ref{in1}) only when $ |\Omega| \gg |\Omega_p| $, i.e. in the central regions far enough from the corotation.
Fig.\,\ref{fig1s} shows the ratio (\ref{ratio}) for the same 3 models as in Fig.\,\ref{fig_shust}. It is seen that $ A \ll 1$ for the slowest mode, so the validity of the theory is justified for such modes. 
The fact that the theory has a certain field of application justifies the main conclusions of the theory (classification of modes, physics of instabilities), derived in Section 3. They can be adequate even beyond the framework of rigorous validity of our approximate theory.
For the other two modes from Fig.\,\ref{fig1s}, the ratio is considerably smaller than 1 only in the central regions. Nevertheless, we can hope that the approach is still valid even for these modes, at least quantitatively, taking into account that the perturbation amplitude decrease with radius, being relatively small at the corotation. Besides, there is one more factor that weakens the role of CR term; it is considered in the next point.

\vspace{12pt}\noindent
3. The coefficient $ a_l $ in the expansion (\ref{dfe1}) is proportional to the linear combination of the derivatives of the unperturbed distribution function 
$ f_0({\bf I}) $ (see below, item 5):
\begin{equation}
a_l \sim f'_{0,lm}({\bf I}),
\end{equation} 
where
\begin{equation}
f'_{0,lm}({\bf I}) = l\frac{\p f_0({\bf I})}{\p I_1} + m\frac{\p f_0({\bf I})}{\p I_2}.
\end{equation} 
Thus, for the bi-symmetric mode $ m=2 $ one should actually calculate the ratio
\begin{equation}
A' = \frac{f'_{0,02}({\bf I})}{|\omega - m\Omega_2|} : 
\frac{f'_{0,-12}({\bf I})}{|\omega-m\Omega_{pr}|},
\label{ratio2}
\end{equation} 
rather than (\ref{ratio}).
In the epicyclic approximation, which is valid everywhere beyond the very central parts of galactic discs:
\begin{equation}
\left| \frac{\p f_0}{\p I_1}\right| \gg 
\left|\frac{\p f_0}{\p I_2}\right|.
\end{equation} 
Fortunately, for the CR term $ l=0 $, so the ratio (\ref{ratio2}) contains the additional small factor (compared to (\ref{ratio})). Apparently, this may be very important for explaining why our theory is valid even for fastest modes. 

\vspace{12pt}\noindent
4. Intuition suggests that higher harmonics (with $ |l|\geq 2 $) cannot considerably influence the large scale modes, which are the subject of our study. The particular number of harmonics, needed for the mode computation, is determined ``experimentally''. The comparison with $ N $-body simulations shows that the only ILR term provides a reasonable accuracy. In the next point we give a more detailed analysis of the relative roles of the terms in the expansion (\ref{dfe1}).

\vspace{12pt}\noindent
5. Let us derive the generalization of the basic integral equation that takes into account all a resonanses of the perturbation with fixed azimuthal number $m$. Let the perturbed potential and the perturbed distribution function are expanded in Fourier series
\begin{equation}
\Phi({\bf I}, w_1) = \sum\limits_l\Phi_l({\bf I}) e^{ilw_1},\qquad
f({\bf I}, w_1) = \sum\limits_lf_l({\bf I}) e^{ilw_1},
\label{pdfexp}
\end{equation} 
where $ {\bf I} = (I_1, I_2) $ are usual radial and azimuthal actions, $ w_1 $ is the radial angular variable. Substituting the full potential and the full distribution function 
\begin{equation}
\Phi_0 + \Phi({\bf I}, w_1)e^{imw_2},\qquad
f_0 + f({\bf I}, w_1)e^{imw_2},
\label{pdffull}
\end{equation} 
into the collisionless Boltzmann equation, one can obtain after linearization
\begin{equation}
f_l(l\Omega_1 + m\Omega_2 - \omega) = \Phi_l f'_{0,lm},
\label{cbe1}
\end{equation} 
where the function $ f'_{0,lm} $ is
\begin{equation}
f'_{0,lm}({\bf I}) = l\frac{\p f_0({\bf I})}{\p I_1} + m\frac{\p f_0({\bf I})}{\p I_2}.
\end{equation} 

The second relation between $\Phi_l({\bf I})$ and $ f_l({\bf I}) $ is provided by the Poisson equation
\begin{equation}
\Phi({\bf I}, w_1)e^{imw_2} = 
-G\int \d {\bf I}' \d {\bf w}' \frac{f({\bf I}', w'_1)e^{imw'_2}}{r_{12}}.
\end{equation} 
Multiplying both sides of the equation by $ e^{imw_1/2} $ and introducing function
\begin{equation}
\varphi_1(w_1) =  w_2 - w_1/2 - \varphi
\end{equation} 
(see Section 2), one can obtain
\begin{equation}
\sum\limits_l \Phi_l({\bf I}) e^{i(l+m/2)w_1} = 
-G\int \d {\bf I}' \d w_1 \sum\limits_{l'} f_{l'}({\bf I}') 
e^{i(l'+m/2)w'_1} e^{im\delta\varphi_1} 
\psi(r,r').
\end{equation}
where $ \delta\varphi_1 = \varphi'_1-\varphi_1 $. 
Denoting $ \Pi_{l,l'} $ the integral
\begin{equation}
\Pi_{l,l'}({\bf I},{\bf I}') = \int\d w_1\d w'_1 \psi(r,r')
e^{i(l'+m/2)w'_1 - i(l+m/2)w_1} e^{im\delta\varphi_1},
\end{equation} 
one obtains the second relation:
\begin{equation}
\Phi_l = -\frac{G}{2\pi}\int \d {\bf I}' \Pi_{l,l'}({\bf I},{\bf I}')
f_{l'}({\bf I}')
\end{equation} 
(here and below the summation over the blind index $ l' $ is assumed).
 Thus, expressing $ f_l $ from (\ref{cbe1}) one can derive the final set of integral equations:
\begin{equation}
\Phi_l ({\bf I}) = 
-\frac{G}{2\pi} \int \d {\bf I}' \Pi_{l,l'}({\bf I},{\bf I}') 
f'_{0,l'm}({\bf I'})
\frac{\Phi_{l'} ({\bf I}')}{l\Omega_1 + m\Omega_2 - \omega}
\label{bs1}
\end{equation}

For numerical calculations, it is more convinient to rewrite the equation (\ref{bs1}) in the form of the classical eigen-value problem. Indeed, using (\ref{cbe1}) one can obtain
\begin{equation}
f_l({\bf I})(l\Omega_1 + m\Omega_2 - \omega) = 
-\frac{G}{2\pi}f'_{0,lm}({\bf I}) \int \d {\bf I}' \Pi_{l,l'}({\bf I},{\bf I}') f_{l'}({\bf I}')
\label{bs2}
\end{equation} 
Using the discretization of the integrals $\int \d{\bf I}' \to \sum \Delta {\bf I}'$, the eigen-frequencies can be obtained  from the condition:
\begin{equation}
\textrm{det}\, \left| \frac{G}{2\pi} f'_{0,lm}({\bf I})
\Pi_{l,l'}({\bf I},{\bf I}')\Delta {\bf I}' + {\bf E}(l\Omega_1({\bf I}) + m\Omega_2({\bf I}) - \omega)\right| =0,
\label{det}
\end{equation} 
where ${\bf E}$ is the unity matrix. Note that the basic integral equation (\ref{ie2}) is the special case of (\ref{bs2}) when $ l=l'=-1 $.

In fact, the equations (\ref{bs2},\,\ref{det}) give a new formulation of the general eigen-value problem for galactic discs. This formulation can be considered as an alternative for the well-known matrix approach of Kalnajs 
(ApJ, v.212, p.637 (1977)). 

Let us employ the equation (\ref{det}) to define more exactly the eigen-frequencies of the bi-symmetric $ m=2 $ modes in the Kalnajs model $ (6,0,1.0,0.25) $, which was already studied in Section 4. Athanassoula \& Sellwood reported that there are two unstable bar-modes: $\omega_1 = 0.465 + 0.066i$, $\omega_2 = 0.33 + 0.058i$. We found that for this model the Lynden-Bell derivative is positive everywhere, so only bar-modes with $ \Omega_p > (\Omega_{pr})_{\max} $ were possible. Using the basic integral equation (\ref{ie2}), we can determine only the pattern speed of the modes; they found to be $ \re\omega_1 = 0.44$, $ \re\omega_2 = 0.35 $. We argued that the growth rate for these modes can be obtained, taking into account the angular momentum exchange at the resonances (primarily, CR and OLR). Rough estimates gave us $ \gamma_1=0.117 $ and $ \gamma_2 = 0.054 $; notably, the OLR contribution were dominating in both cases.

Restricting ourselves to consideration of three terms with $ |l|\leq 1 $ in the set of equations (\ref{bs2}), we can study the impact of three main resonances: ILR, CR, and OLR. First of all, we obtained the unstable bar-modes, when all three terms are present (run 1). Then, the influence of a particular resonance can be ``measured'' separately by omitting an unnecessary term. In such a way, we study the influence of CR term, switching off the OLR term (run 2). After that we study the impact of the OLR term, omitting CR term (run 3). 
In all these three cases, when the principal ILR term is present, the pictures of the unstable modes in the complex $ \omega $-plane resemble each other. The situation has changed when we dropped the ILR term. In the result we obtained unrealistic picture of a sea of unstable modes with growth rates less than $7\cdot 10^{-4}$. This fact certainly confirms the dominating role of the ILR term.

The computed eigen-frequencies are given in the table.

\vspace{2mm}
\begin{center}
\begin{tabular}{|c|l|c|c|c|}
\hline Run &Resonances & Mode 1 & Mode 2 \\ 
\hline 1&ILR, CR, OLR & $0.48+0.058i$ & $0.38+0.024i$ \\ 
\hline 2&ILR, CR & $0.43+0.015i$ & --- \\ 
\hline 3&ILR, OLR & $0.49+0.036i$ & $0.38+0.024i$ \\ 
\hline 
\end{tabular} 
\end{center}

\vspace{2mm}
It can be seen that:
\begin{itemize}
\item All pattern speeds gained small positive biases, compared to the pattern speeds obtained from the basic integral equation (\ref{ie2}).

\item When all three terms are taken into account, both the pattern speed and the growth rate of the most unstable mode (mode 1) is determined with very good accuracy. Separate consideration of CR and OLR (runs 2 and 3) shows that the contribution of OLR term into the growth rate exceeds one of CR.  

\item From comparing the eigen-frequency of mode 2 obtained in runs 1 and 3, it follows that mode's growth is entirely due to the interaction of the gravitational potential of the mode with the resonance stars at OLR.

\end{itemize} 

%\vspace{2mm}
In principle, equations (\ref{bs2}) allow to consider any number of terms, although it can be checked that the terms with $ |l|\geq 2 $ are of less importance. For example, taking account of terms $l = \pm 2$ will not lead to any new unstable modes. The fastest modes become rather more unstable: $ \gamma_1 =0.08 $, $ \gamma_2= 0.032$.

\end{document}